\documentclass[aps,prd,superscriptaddress,preprintnumbers,nofootinbib,notitlepage]{revtex4-2}
\usepackage[latin9]{inputenc}
\usepackage{bm}
\usepackage{amsmath}
\usepackage{amsthm}
\usepackage{amssymb}
\usepackage[colorlinks=true]{hyperref}

\newcommand{\dd}{\mathrm{d}}

\begin{document}

\preprint{YITP-19-75, IPMU19-0106}

\title{On the absence of conformally flat slicings of the Kerr spacetime}

\author{Antonio De Felice}
\email{antonio.defelice@yukawa.kyoto-u.ac.jp}
\affiliation{Center for Gravitational Physics, Yukawa Institute for Theoretical Physics, Kyoto University, 606-8502, Kyoto, Japan}
\author{Fran\c{c}ois Larrouturou}
\email{francois.larrouturou@iap.fr}
\affiliation{Institut d'Astrophysique de Paris, UMR 7095, CNRS, Sorbonne Universit{\'e},\\ 98\textsuperscript{bis} boulevard Arago, 75014 Paris, France}
\author{Shinji Mukohyama}
\email{shinji.mukohyama@yukawa.kyoto-u.ac.jp}
\affiliation{Center for Gravitational Physics, Yukawa Institute for Theoretical Physics, Kyoto University, 606-8502, Kyoto, Japan}
\affiliation{Kavli Institute for the Physics and Mathematics of the Universe (WPI), The University of Tokyo Institutes for Advanced Study, The University of Tokyo, Kashiwa, Chiba 277-8583, Japan}
\author{Michele Oliosi}
\email{michele.oliosi@yukawa.kyoto-u.ac.jp}
\affiliation{Center for Gravitational Physics, Yukawa Institute for Theoretical Physics, Kyoto University, 606-8502, Kyoto, Japan}

\date{\today}

\begin{abstract}
This work investigates the possibility of achieving a conformally flat slicing of the Kerr spacetime. 
 We consider a hypersurface of the form $t = F(r,\theta,a)$, where $(t,r,\theta,\phi)$ are the Boyer-Lindquist coordinates, solve for a vanishing Cotton-York tensor of the induced metric order by order in the spin parameter $a$, and show that the procedure fails at the fifth order.
We also prove that no coordinate change can induce a spatially flat recasting of the Kerr(-de Sitter) metric, beyond linear order in $a$, adopting a more general ansatz depending on $\phi$. 
\end{abstract}

\maketitle

\section{Introduction and motivations} 
\label{sec:intro}

From their theoretical discovery by Karl Schwarzschild in 1916~\cite{Schwarzschild_OP} to the first direct observation of their immediate vicinity more than a century later~\cite{EHT_M87}, black holes (BHs) have been one of the cornerstones of modern gravitational physics. While a static BH is one of the simplest objects in General Relativity (GR) at least from a theoretical point of view, it is more difficult to study spinning BH spacetimes: their construction was achieved only in 1963~\cite{Kerr_OP}. Those objects possess rich phenomenology: for example, they can trigger the so-called ``Active Galactic Nuclei'', induce Penrose processes or even, if in binary or triplets, emit detectable gravitational radiation.\\

The detection of gravitational radiation relies heavily on Numerical Relativity, notably to estimate the gravitational waveform produced during the merger phase (which is the dominant part of the signal for the binary BH induced events detected by the LIGO/Virgo collaboration~\cite{LVC_GWTC1}).
In order to find initial data for these numerical studies, it has been common to rely on conformal flatness, \emph{i.e.}~that the spatial metric induced by such foliations can be written as $\gamma_{ij} = \Omega^2(r,\theta,\phi)\, \eta_{ij}$, where $\Omega$ is a free function of the spatial coordinates and $\eta$ is any usual flat metric. Classical examples using or being simplified by this assumption include the Misner~\cite{Misner} or Bowen-York~\cite{BowenYork} initial data, the Isenberg-Wilson-Mathews formulation~\cite{Isenberg,WilsonMathews}
or the puncture framework~\cite{BrandtBrugmann}, all of which have been widely used in the literature\footnote{For a review, see \cite{Cook} or \cite{Gourgoulhon}. Conformally flat initial data has notably been used for the first promising simulation of a binary black hole space-time \cite{Pretorius}. For more recent work using conformal flatness see for example \cite{Kyutoku,VanoVinuales}.}.
In this context, it has been therefore natural to seek for conformally flat foliations of the different BH spacetimes.\\

Such a conformally flat slicing is trivially realised in the case of static BHs. A static and spatially flat BH solution was found as early as in 1921, independently by P. Painlev\'{e}~\cite{Painleve} and A. Gullstrand~\cite{Gullstrand}. This solution was recognised as a simple coordinate transformation of the usual Schwarzschild coordinate system by G.~Lema\^{i}tre, 12 years later~\cite{Lemaitre_PG}.
In the case of rotating BHs, the game is more involved and a first approach, conducted by A.~Garat and R.~H.~Price~\cite{Garat_Price}, ended up with a no-go result indicating that the Kerr metric does not allow for a conformally flat slicing\footnote{Conformally flat initial data have nevertheless been commonly used, in particular for slowly spinning black holes. In such a case, one has to put up with spurious ``junk'' radiation (see \emph{e.g.}~\cite{CookYork,Gleiser,DainLoustoTakahashi,Lovelace2008}). See also \cite{KrivianPrice,Hannam,Lovelace2008,Lovelace2011} for binary black hole initial data beyond conformal flatness.}.
However, as discussed in the next section, their no-go result is based on two restrictive assumptions. The main purpose of the present paper is to strengthen the no-go result by relaxing one of their assumptions. \\

The present work has significant implications also in a different context. In~\cite{MTMG_BH} the authors have proven that any solution of GR that admits a spatially flat slicing is also a solution of an alternative theory of gravity, namely the \emph{Minimal Theory of Massive Gravity} (MTMG)~\cite{MTMG_OP,MTMG_Pheno}. So it has been shown that MTMG admits static BH solutions, and the next step is naturally to investigate whether it also admits rotating BH solutions. For this reason, it is of physical interest to elucidate whether the Kerr(-de Sitter) spacetime admits spatially flat slicings, which are a subclass of conformally flat slicings, or not. However, while the Schwarzschild solution of MTMG is in the Painlev\'{e}-Gullstrand slicing, the previous no-go result of \cite{Garat_Price} applies only to those slicings that reduce to the Schwarzschild slicing in the non-spinning limit. The extended no-go result of the present paper is general enough to exclude a conformally flat slicing that in the non-spinning limit reduces to the Painlev\'{e}-Gullstrand slicing. A full proof of the no-go result for spatially flat slicings of the Kerr(-de Sitter) spacetime (\emph{i.e.}~via a general change of coordinates) is separately presented in the Appendix of this work. The new no-go results shown in the present paper imply that rotating BHs solutions cannot be implemented in MTMG by the aforementioned procedure. This implies either that MTMG accommodates the Kerr solution in a different way or that rotating BHs should deviate from the Kerr spacetime. In the latter case BHs would provide a window to distinguish MTMG and GR observationally. \\

This work is organised as follows: in Sec.~\ref{sec:strategy} we expose our strategy to construct a conformally flat slicing of the Kerr metric, and we apply it in Sec.~\ref{sec:cancel}, up to its failure at the fifth order in the spin parameter $a$. We then discuss this no-go result an conclude in~\ref{sec:concl}. As we were initially interested in spatially flat slicings, we also set a no-go result on the construction of such foliations by a general coordinate change in App.~\ref{app:flat}. Any lengthy expressions we had to deal with are presented in App.~\ref{app:lengthy_expr}.

\section{Strategy}\label{sec:strategy}

We write the Kerr line element in Boyer-Lindquist coordinates $(t,r,\theta,\phi)$~\cite{Boyer_Lindquist}
\begin{equation}\label{eq:KerrBL}
\dd s^2 = 
-\frac{\Delta-a^2\sin^2\theta}{\Sigma}\,\dd t^2 
+ \frac{\Sigma}{\Delta}\,\dd r^2 
+ \Sigma\,\dd\theta^2 + \frac{\left(r^2+a^2\right)^2
-a^2\Delta \sin^2\theta}{\Sigma}\,\sin^2\theta \dd\phi^2
-\frac{4aMr\sin^2\theta}{\Sigma}\,\dd t\dd \phi,
\end{equation}
with the auxiliary functions
\begin{equation}\label{eq:KerrSigmaLambda}
    \Delta = r^2-2 M r+a^2
    \qquad \text{and} \qquad 
    \Sigma = r^2 + a^2\cos^2\theta.
\end{equation}

In order to seek for a possible conformally flat slicing, we introduce a hypersurface~\footnote{If one finds a conformally flat hypersurface of this form, then one can easily promote it to a conformally flat foliation, \emph{i.e.}~a family of conformally flat hypersurfaces, by simply adding different constants to $f_0(r)$. This procedure is guaranteed to work since the spacetime is invariant under a constant shift of $t$.} specified by $t = F(r,\theta,a)$ while imposing $F(r,\theta,0) = f_0(r)$. Such form is motivated by the original symmetries of the Kerr metric: no azimuthal dependency is included so as to preserve axisymmetry and, in the non-spinning limit, it reduces to the Schwarzschild one, that is a spherically symmetric form. Note that this slicing is more general than the one used in~\cite{Garat_Price}, in which $F(r,\theta,0) = const.$~was imposed. This restriction\footnote{They also prevent $\phi$ dependencies as, to quote their own words, \emph{Gaining the advantages of conformal flatness while losing axisymmetry would be a Pyrrhic victory}.} reduces drastically the field of possible solutions: it notably prevents solutions that would reduce to the Painlev\'{e}-Gullstrand (\emph{i.e.}\ spatially flat) coordinates in the limit $a =0$. On the other hand, in this work, by assuming $F(r,\theta,0) = f_0(r)$, we keep the possibility of finding a slicing that in the static limit reduces to a non-Schwarzschild slicing, including the Painlev{\'e}-Gullstrand one.\\

Starting from a general $a$-expanded slicing
\begin{equation}
t = f_0(r) + \sum_{n=1}^\infty a^nf_n(r,\theta),
\end{equation}
we will compute the Cotton-York tensor on the induced spatial hyper-surfaces, given by
\begin{equation}
\mathcal{C}^i_j = \varepsilon^{ikl} \nabla_k\left(R_{jl}-\frac{1}{4}R \,\gamma_{jl}\right)\!,
\end{equation}
where $\nabla_i$, $R_{ij}$ and $R$ are respectively the covariant derivative, the Ricci tensor and the Ricci scalar associated with the induced 3D metric $\gamma_{ij}$, and $\varepsilon^{ijk}$ is the Levi-Civita tensor. As there is a one-to-one correspondence between conformal flatness and the cancellation of the Cotton-York tensor in 3-dimensions, we will try to solve the equation $\mathcal{C}^i_j=0$ order by order in $a$, and thus constrain the functions $f_n(r,\theta)$.

\section{Proof of non cancellation of the Cotton-York tensor}\label{sec:cancel}

The great advantage of our assumption $F(r,\theta,0) = f_0(r)$ is that the Cotton-York tensor automatically vanishes at the zeroth order in $a$. This is naturally linked to the fact that the Schwarzschild metric is conformally flat.\\

At the linear order in $a$, the $r\phi$-component of the Cotton-York tensor is factorisable as
\begin{equation}\label{eq:CY_lin}
\mathcal{C}^r_\phi = a\,\frac{\mathcal{A}_1(r)\, \mathcal{B}_1(r,\theta)}{2r\sin^4\theta\left(r^2-(r-2M)^2(f_0')^2\right)} + \mathcal{O}(a^2),
\end{equation}
with
\begin{subequations}
\begin{align}
& \mathcal{A}_1 = r\left(1-\frac{2M}{r}\right)f_0''+\left(1-\frac{2M}{r}\right)^2\left(1-\frac{3M}{r}\right)(f_0')^3-\left(1-\frac{5M}{r}\right)f_0',\label{eq:CY_lin_coeff_A}\\
& \mathcal{B}_1 = \frac{\partial_\theta^3 f_1+\cot\theta\,\partial_\theta^2f_1+(1-\cot^2\theta)\partial_\theta f_1}{\sin\theta}.\label{eq:CY_lin_coeff_B1}
\end{align}
\end{subequations}

So at this level, it is clear that the possible solutions split in two branches:\footnote{The two branches still exist for a parameterization $F(r,\theta,\phi,a)$ with $F(x^i,0) = f_0(r)$. However, in this case, the $\phi$ dependence drastically complicates the resolution of the first branch by introducing highly non-linear PDEs. Nevertheless, the whole argument of non-existence of conformally flat slicing is not significantly changed, for the second branch, by this additional dependence in $\phi$.}  in the first one, $f_0$ has to be constrained so that $\mathcal{A}_1$ vanishes whereas in the second one, we have to solve $\mathcal{B}_1 = 0$ \emph{w.r.t.}~$f_1$. 

\subsection{First branch}

\paragraph{First order:} Solving $\mathcal{A}_1(r) =0$ in terms of $f_0'$ yields
\begin{equation}\label{eq:Br1_f0}
f_0'(r) = \pm \frac{r^{5/2}}{\left(r-2M\right)\sqrt{r^3+\lambda\left(r-2M\right)}},
\end{equation}
where $\lambda$ is an integration constant with dimension of squared mass.
Note that we present here $f_0'$ and not $f_0$ as the latter will never appear in the equations due to the time shift symmetry. 
When this solution is injected, the whole Cotton-York tensor vanishes at the first order in $a$.\\

\paragraph{Second order:} At the second order in $a$, the $rr$-component of the Cotton-York tensor reads
\begin{equation}
\mathcal{C}^r_r = -\frac{3Ma^2\sin\theta}{r^4}\left[\partial_\theta^3f_1+5\cot\theta\,\partial_\theta^2f_1+\frac{10\cos^2\theta-3}{\sin^2\theta}\,\partial_\theta f_1\right]\!,
\end{equation}
which, imposing regularity on $]0,2\pi[$, yields $f_1=\bar{f}_1(r)+  f_{1,s}(r)/\sin^2\theta$. Injecting it in the $r\theta$-component, it comes
\begin{equation}
\mathcal{C}^r_\theta = -\frac{6Ma^2}{r^2(r^3+\lambda(r-2M))}\left[\left(2+\frac{3\lambda}{r^2}-\frac{7M\lambda}{r^3}\right)f_{1,s}-r\left(1+\frac{\lambda}{r^2}-\frac{2M\lambda}{r^3}\right)f_{1,s}'\right] + \mathcal{O}(a^3),
\end{equation}
which imposes $f_{1,s} = \lambda_1 r^{7/2}/\sqrt{r^3 +\lambda(r-2M)}$, with $\lambda_1$ a constant with dimension of inverse squared mass. One can then express the $\theta\phi$-component as
\begin{equation}
\mathcal{C}^\theta_\phi = \frac{a^2}{\lambda^2r^2\sin^{10}\theta\left(r^3+\lambda(r-2M)\right)} \sum_{n=0}^{5} \Gamma_n(r)\,\cos^{2n}\theta + \mathcal{O}(a^3).
\end{equation}
Here, $\Gamma_n$ ($n=0,\cdots,5$) are functions of $r$, which we do not need to fully specify for the argument. As $\mathcal{C}^\theta_\phi$ should vanish for any $\theta$, each of the $ \Gamma_n(r)$ should vanish by its own. However, we have
\begin{equation}
\Gamma_5(r) = M\lambda\left\lbrace 42M\lambda^2\left(r-2M\right)-3\lambda r^3\left(r-16M\right) -3r^6\right\rbrace\!.
\end{equation}
One can not simply impose $\lambda\rightarrow 0$ as $\mathcal{C}^\theta_\phi$ would blow up, due to the presence of $\lambda^2$ in the denominator, so it is impossible to cancel $\mathcal{C}^i_j$ at the second order in $a$, if $f_0$ is of the form~\eqref{eq:Br1_f0}.

\subsection{Second branch}\label{sec:cancel_Br2}

\paragraph{First order:} In this branch we have to solve $\mathcal{B}_{1} =0$ in terms of $f_1$, which yields simply 
\begin{equation}
f_1(r,\theta) = \bar{f}_1(r) + f_{1,c}(r)\cos\theta\,.
\end{equation}
The slicing at the first order in $a$ is hence parameterized by two undetermined functions of $r$, $\{\bar{f}_1,f_{1,c}\}$. Note that $\bar{f}_1(r)$ can be reabsorbed in $f_0(r)$.
When this solution is injected, the whole Cotton-York tensor vanishes at the first order in $a$.\\

\paragraph{Second order:} The $rr$-component of the Cotton-York tensor reads
\begin{equation}
\mathcal{C}^r_r = \frac{4Ma^2\left(3\cos^2\theta-1\right)}{r^4}\,f_{1,c}(r) + \mathcal{O}(a^3),
\end{equation}
which imposes $f_{1,c}=0$: we are left with only $f_1 = \bar{f}_1(r)$ that does not appear in $\mathcal{C}^i_j$ at this order. The next non-vanishing component is
\begin{equation}
\mathcal{C}^r_\phi = -\frac{a^2}{2r^3}\,\frac{\mathcal{A}_1(r)\mathcal{B}_2(r,\theta)+4\cos\theta\mathcal{S}_2(r)}{1 -\left(1-\frac{2M}{r}\right)^2(f_0')^2} + \mathcal{O}(a^3)\,,\label{eq:crphi_b2_a2}
\end{equation}
where $\mathcal{A}_1$ is still given by \eqref{eq:CY_lin_coeff_A}, $\mathcal{B}_{2}$ is given by \eqref{eq:CY_lin_coeff_B1}, when substituting $f_2$ to $f_1$, and where the expression of the source term $\mathcal{S}_{2}$ is given in App.~\ref{app:lengthy_expr}. Solving for $f_{2}$, it finally comes
\begin{equation}
F(r,\theta,a) = f_0(r) + a\,\bar{f}_1(r)
 + a^2\left[ \bar{f}_{2}(r) + f_{2,c}(r)\cos\theta  + \hat{f}_2(r)\cos^2\theta \right]
 + \mathcal{O}(a^3),
\end{equation}
where $\{f_0,\bar{f}_1,\bar{f}_2,f_{2,c}\}$ are four free functions and $\hat{f}_2(r)$ is given in terms of $f_0'$ and $f_0''$ in App.~\ref{app:lengthy_expr}.\\

\paragraph{Third order: } Again, the $rr$-component of the Cotton-York tensor imposes a first set of conditions on the lower-order functions.
To let $\frac{d^3}{d(\cos\theta)^3}\mathcal{C}_r^r$ vanish, one must impose
\begin{equation}
63\frac{M^2}{r^2}-f_0'^2\left(1-\frac{2M}{r}\right)^2\left(5+\frac{2M}{r}+\frac{45M^2}{r^2}\right) + f_0'^4\left(1-\frac{2M}{r}\right)^4\left(5 + \frac{12M}{r}\right)+rf_0'f_0''\left(1-\frac{2M}{r}\right)^3\left(5+\frac{9M}{r}\right)=0\,,\label{eq:diffeq_f0}
\end{equation}
which can be integrated to give $f_0(r) = \hat{f}_0(r)$. Here, the explicit expression of $\hat{f}_0(r)$ is shown in App.~\ref{app:lengthy_expr}. In order to keep a compact notation we will keep $\hat{f}_0$ as a shorthand in most expressions. This also fully fixes the form of $\hat{f}_2(r)$. With $f_0$ fixed, 
\begin{equation}
\mathcal{C}_r^r = -\frac{12 a^3M}{r^4}(3\cos^2\theta-1)f_{2,c} + \mathcal{O}(a^4) = 0\,,
\end{equation}
gives $f_{2,c} = 0$. Focusing on the $\mathcal{C}^r_\phi$ component then gives
\begin{equation}
   \frac{9M}{2r}\frac{\frac{7 M}{r} - \hat{f}_0'^2\left(1-\frac{2M}{r}\right)^2\left(2+\frac{3M}{r}\right)}{1-\frac{2M}{r}}\,\mathcal{B}_3(r,\theta) + \frac{2\cos\theta}{3r}\mathcal{S}_3(r) = 0\,, \label{eq:diffeq_f3}
\end{equation}
where $\mathcal{B}_{3}$ is given by \eqref{eq:CY_lin_coeff_B1}, when substituting $f_3$ to $f_1$, and where the expression of the source term $\mathcal{S}_{3}$ is given in App.~\ref{app:lengthy_expr}. This allows one to fix
\begin{equation}
f_3(r,\theta) = \bar{f}_{3}(r) + f_{3,c}(r) \cos\theta + \hat{f}_{3}(r) \cos^2\theta\,,
\end{equation}
with $\hat{f}_{3}(r)$ given in terms of $\bar{f}_1(r)$, which leads to 
\begin{equation}
F(r,\theta,a) = \hat{f}_0(r) + a\,\bar{f}_1(r)
 + a^2\left[ \bar{f}_{2}(r) + \frac{\hat{f}_0'}{3r}\left(1-\frac{2M}{r}\right)\cos^2\theta\right]
 +a^3\left[\bar{f}_{3}(r) + f_{3,c}(r) \cos\theta + \hat{f}_{3}(r) \cos^2\theta\right]+ \mathcal{O}(a^4).
\end{equation}
\paragraph{Fourth order: } Demanding $\frac{d^3}{d(\cos\theta)^3}\mathcal{C}_r^r$ to vanish, one finds the equation
\begin{equation}
\bar{f}''_1 -\bar{f}_1'\frac{63\frac{M^2}{r^2}-3\hat{f}_0'^4\left(1-\frac{2M}{r}\right)^4\left(5 + \frac{12M}{r}\right)+\hat{f}_0'^2\left(1-\frac{2M}{r}\right)^2\left(5+\frac{2M}{r}+\frac{45M^2}{r^2}\right)}{\hat{f}_0'^2\left(1-\frac{2M}{r}\right)^3\left(5+\frac{9M}{r}\right)r} = 0\,,\label{eq:diffeq_f1b}
\end{equation}
which can be solved by $\bar{f}_1(r) = \hat{f}_1(r)$ given in App.~\ref{app:lengthy_expr}. Plugging this solution back and demanding the vanishing of $\mathcal{C}_r^r$, one finds that $f_{3,c} = 0$. The vanishing of $\mathcal{C}^r_\phi$ then translates into a differential equation for $f_4$,
\begin{equation}
\frac{9M}{2r}\left[\frac{7M}{r} - \left(1-\frac{2M}{r}\right)^2\left(2+\frac{3M}{r}\right)\hat{f}_0'^2\right]\mathcal{B}_4(r,\theta) + \frac{\cos\theta\,\mathcal{S}_{4,1}(r)+ \cos^3\theta\,\mathcal{S}_{4,3}(r)}{3\left(1-\frac{2M}{r}\right)\left(5+\frac{9M}{r}\right)\left[\hat{f}_0'^2\left(1-\frac{2M}{r}\right)^2-1\right]\hat{f}_0'r^3}=0\,,\label{eq:diffeq_f4}
\end{equation}
where $\mathcal{B}_{4}$ is given by \eqref{eq:CY_lin_coeff_B1}, when substituting $f_4$ to $f_1$, and where the source terms $\mathcal{S}_{4,2}$ and $\mathcal{S}_{4,4}$ are given explicitly in App.~\ref{app:lengthy_expr}. Therefore, we obtain the full solution
\begin{equation}
f_4(r,\theta) = \bar{f}_{4}(r) + f_{4,c}(r) \cos\theta + \hat{f}_{4,2}(r) \cos^2\theta + \hat{f}_{4,4}(r) \cos^4\theta\,,
\end{equation}
where $\hat{f}_{4,2}(r)$ and $\hat{f}_{4,4}(r)$ are fully dependent functions of $r$ and can for example be given in terms of $\hat{f}_0$, $\hat{f}_1$, and $\bar{f}_2$. At the fourth order we thus have
\begin{equation}
\begin{aligned}
F(r,\theta,a) = \hat{f}_0(r) + a\,\hat{f}_1(r)
 + &a^2\left[ \bar{f}_{2}(r) + \frac{\hat{f}_0'}{3r}\left(1-\frac{2M}{r}\right)\cos^2\theta\right]
 +a^3\left[\bar{f}_{3}(r) + \frac{\hat{f}_1'}{3r}\left(1-\frac{2M}{r}\right) \cos^2\theta\right]\\ 
 &+a^4\left[\bar{f}_{4}(r) + f_{4,c}(r) \cos\theta + \hat{f}_{4,2}(r) \cos^2\theta + \hat{f}_{4,4}(r) \cos^4\theta \right] + \mathcal{O}(a^5)\,,
 \end{aligned}
\end{equation}
which lets the Cotton-York tensor completely vanish at this order, and which depends on four free functions $\{\bar{f}_2,\bar{f}_3,\bar{f}_4,f_{4,c}\}$ and four integration constants (within $\hat{f}_0$ and $\hat{f}_1$).
\paragraph{Fifth order: } Finally, by considering
\begin{equation}
\begin{aligned}
\frac{d^5}{d(\cos\theta)^5}\mathcal{C}_r^r = -\frac{7 a^5}{54 r^6 \left(5+\frac{9M}{r}\right)^3 }\frac{\mathcal{N}_1(r)\mathcal{N}_2(r)}{\mathcal{D}_1(r)(\mathcal{D}_2(r))^{1/2}} + \mathcal{O}(a^6)\,,\label{eq:Crr5_d5z}
\end{aligned}
\end{equation}
with the fully (up to an integration constant) determined $\mathcal{N}_1$, $\mathcal{N}_2$, $\mathcal{D}_1$, $\mathcal{D}_2$ given in App.~\ref{app:lengthy_expr}, one finds that the Cotton-York tensor cannot identically vanish under the assumptions made above. This concludes our argument.

\section{Conclusion and discussion}
\label{sec:concl}

Conformally flat slicings are of prime importance when dealing with realistic spacetimes, as they allow to efficiently obtain initial data for numerical computations.
While the conformally flat slicings of static BHs have been known for a long time, it has been impossible to find their equivalent for rotating BHs.
A.~Garat and R.~H.~Price showed that no slicing of the form $t = F(r,\theta,a)$ with $F(r,\theta,0) = const.$ could support conformal flatness~\cite{Garat_Price}.
In this work, we followed their steps and relaxed one of their assumptions by taking $t = F(r,\theta,a)$ with $F(r,\theta,0) = f_0(r)$, which has notably the advantage of including the Painlev{\'e}-Gullstrand coordinate change. 
Even under this weaker restriction, we have demonstrated that it is not possible to find conformally flat hypersurfaces.
The next steps would naturally be to either relax the assumption that $F(r,\theta,0) = f_0(r)$ or/and examine a parameterization $F(r,\theta,\phi,a)$~\footnote{A straightforward analysis shows that this possibility fails at the fifth order for the second branch. Indeed, the structure of the argument doesn't change and the same component $\mathcal{C}^r_r$, at the fifth-order in $a$, can be used to conclude the argument.}. These extensions could then potentially yield a stronger no-go result for the construction of a conformally flat slicing of the Kerr spacetime, but at the cost of hiding its original symmetries. Finally, note that our results agree, and may be possibly further connected, with the findings of \cite{Kroon} in an expansion at infinity.\\

The non-existence of conformally flat slicings of the Kerr spacetime can be linked to the failure of mimicking the exterior of a Kerr BH with ordinary matter.
In general relativity, a well-known theorem due to Jebsen \cite{Jebsen} and Birkhoff \cite{Birkhoff} states that the exterior solution of all spherically symmetric matter content is the Schwarzschild one.
But no such theorem exists in the spinning case, and usually the multipole moments created by a system of spinning ordinary matter will only asymptotically agree with the Kerr ones.
Nevertheless, some attempts to recreate an external Kerr geometry with matter were made, but they always involve exotic matter (see \emph{e.g.}\ \cite{deFelice_etal,McManus,Poisson}).
But, when dealing with gravitational radiation for an ordinary spinning matter system, the metric perturbation around a Minkowskian background is usually gauged as
\begin{equation}
    h_{00} \simeq \frac{2M}{r}, \qquad 
    h_{0i} \simeq \frac{\epsilon_{ijk}S^jx^k}{r ^3}, \qquad
    h_{ij} \simeq \Omega^2\delta_{ij} + h^\text{rad}_{ij},
\end{equation}
with $S^i$ the spin vector. One can naturally see that for $h^\text{rad}_{ij} =0$ this perturbation is conformally flat.
Thus the fact that the Kerr geometry cannot be conformally flat sliced is in agreement with the fact there is \emph{a priori} no ordinary matter system generating it.\\

The present work was originally motivated by the study of black hole solutions in the minimal theory of massive gravity (MTMG). Indeed, this theory has been shown to admit as solutions all spatially-flat general relativistic spacetimes. The question of the existence of a flat slicing of the Kerr solution, which we show not to exist in App.~\ref{app:flat}, was a motivation to search more generically for conformally flat slicings. The present work leaves open the possibility that, within MTMG, rotating black hole solutions be found in a completely different fashion.

\acknowledgments

The authors would like to thank E.~Gourgoulhon for pointing out the existence of the work done by A.~Garat and  R.~H.~Price~\citep{Garat_Price}, and L.~Blanchet for inspiring discussions about the links between Kerr geometry and ordinary matter and how to solve highly non-linear PDEs.
FL would like to express his gratitude to the Yukawa Institute for Theoretical Physics for hosting him during two weeks of rich and fruitful discussions.
The work of SM was supported by Japan Society for the Promotion of Science Grants-in-Aid for Scientific Research No. 17H02890, No. 17H06359, and by World Premier International Research Center Initiative, MEXT, Japan. MO acknowledges the support from the Japanese Government (MEXT) Scholarship for Research Students.


\appendix

\section{Absence of flat slicings in Kerr-de Sitter spactime}
\label{app:flat}

In this Appendix we demonstrate that there are no spatially flat slicings of the Kerr-de Sitter spactime. We will proceed by performing a general coordinate change. This is equivalent to adopting an ansatz that is more general than the one in the main text used for the proof of the absence of conformally flat slicings.

\subsection{Strategy}

Let's recall the Kerr-de Sitter line element written in Boyer-Lindquist-like coordinates
\begin{equation}
\begin{aligned}
\dd s^2 = \bar{g}_{\mu\nu} \dd x^\mu \dd x^\nu = & 
-\frac{\tilde\Delta-a^2\zeta\sin^2\theta}{\Xi}\,\dd t^2 
+ \frac{\Sigma}{\tilde\Delta}\,\dd r^2 
+ \frac{\Sigma}{\zeta}\,\dd\theta^2 
+ \frac{\left(r^2+a^2\right)^2\zeta-a^2\tilde\Delta \sin^2\theta}{\Xi}\,\sin^2\theta \dd\phi^2\\
& \quad
-\frac{2a\left[6Mr-\Lambda(r^2+a^2)\Sigma\right]}{3\,\Xi}\,\dd t\dd \phi,
\end{aligned}
\end{equation}
where $\Lambda$ is the cosmological constant and
\begin{equation}
\tilde\Delta = \left(r^2+a^2\right)\left(1+\frac{\Lambda r^2}{3}\right)-2 M r,
\quad
\Sigma = r^2 + a^2\cos^2\theta,
\quad
\Xi = \Sigma\,\left(1-\frac{\Lambda a^2}{3}\right)^2,
\quad \text{and} \quad 
\zeta = 1-\frac{\Lambda a^2\,\cos^2\theta}{3}.
\end{equation}
This metric reduces to~\eqref{eq:KerrBL} in the $\Lambda = 0$ case.
We will hereafter perform a general coordinate change $x^\mu \rightarrow \chi^\mu(a,x^\nu) = \{\tau,\rho,\vartheta,\varphi\}$ and ask that the spatially induced metric be flat
\begin{equation}\label{eq:App_aim}
\gamma_{ij} \equiv  \bar{g}_{\mu\nu}\frac{\dd x^\mu}{\dd \chi^i} \frac{\dd x^\nu}{\dd \chi^j}  = \hat\delta_{ij},
\end{equation}
where $\hat\delta$ is the usual three-dimensional Euclidean metric~\footnote{Starting from a rotating spacetime, it is natural to aim for a Minkowskian metric written in Born coordinates, which has a purely flat spatial sector.}
\begin{equation}
\hat\delta_{ij}\dd\chi^i\dd\chi^j = \dd\rho^2+ \rho^2\,\dd\vartheta^2 +\rho^2\sin^2\vartheta\,\dd\varphi^2.
\end{equation}
We will also require the change of coordinates to be invertible, namely that the Jacobian of the transformation is non-vanishing
\begin{equation}\label{eq:App_Jacobian}
\mathcal{J} \equiv \left\vert \frac{\partial \chi^\alpha}{\partial x^\mu} \right\vert \neq 0.
\end{equation}\\

Note that we already know the result when $a=0$: in this case the Boyer-Lindquist-like coordinates reduce to the Schwarzschild-de Sitter ones and thus the transformation to apply is an extended Painlev\'{e}-Gullstrand one (see e.g.~\cite{MTMG_BH})
\begin{equation}
t = \tau + \int^\rho \! \dd u\, \frac{\sqrt{2u\,\mu(u)}}{u-2\,\mu(u)},
\qquad x^i = \chi^i,
\end{equation}
where the effective mass is given by $\mu(r) = M- \Lambda r^3/6$.
Starting from this zeroth-order solution, we expand the coordinate change in $a$ as 
\begin{subequations}
\begin{align}
& t = \tau + \int^\rho \! \dd u\, \frac{\sqrt{2\mu \, u}}{u-2\mu} + \sum_{n=1}^\infty a^n \, T^{(n)}(\rho,\vartheta,\varphi), \\
& r = \rho +\sum_{n=1}^\infty  a^n \, R^{(n)}(\rho,\vartheta,\varphi), \\
& \theta = \vartheta + \sum_{n=1}^\infty  a^n \,\Theta^{(n)}(\rho,\vartheta,\varphi),\\
& \phi = \varphi  + \sum_{n=1}^\infty  a^n \,\Phi^{(n)}(\rho,\vartheta,\varphi),
\end{align}
\end{subequations}
and solve order by order the equation $\mathcal{E}_{ij} \equiv \gamma_{ij} - \hat{\delta}_{ij} = \sum_{n=1}^{\infty}a^n\mathcal{E}^{(n)}_{ij}=0$.
Denoting $F^{(n)}_\mu$ the collection $\{T^{(n)},R^{(n)},\Theta^{(n)},\Phi^{(n)}\}$, one can decompose at any order $\mathcal{E}^{(n)}_{ij} = \mathcal{O}_{ij}\left[F^{(n)}_\mu\right] + \mathcal{S}^{(n)}_{ij}$, where $\mathcal{S}^{(n)}_{ij}$ is a source term (depending only on $F^{(m)}_\mu$ with $1\leq m\leq n-1$) and the linear operator $\mathcal{O}_{ij}\left[F^{(n)}\right]$ is given by
\begin{equation}
\mathcal{O}_{ij}\left[F^{(n)}_\mu\right] 
= \left(g^\text{SdS}_{\mu\nu}\frac{\partial F_\nu^{(0)}}{\partial\chi^i}\frac{\partial}{\partial\chi^j}
+g^\text{SdS}_{\mu\nu}\frac{\partial F_\nu^{(0)}}{\partial\chi^j}\frac{\partial}{\partial\chi^i}
+\frac{\partial g^\text{SdS}_{\nu\lambda}}{\partial x^\mu}\frac{\partial F_\nu^{(0)}}{\partial\chi^i}\frac{\partial F_\lambda^{(0)}}{\partial\chi^j}\right) F_\mu^{(n)},
\end{equation}
where $g^\text{SdS}_{\mu\nu}$ is the usual Schwarzschild-de Sitter metric. Explicitly, it reads
\begin{subequations}\label{eq:App_Oij}
\begin{align}
& \mathcal{O}_{\rho\rho} =  2 \left(\partial_\rho A +\frac{3M+\Lambda\rho^3}{6M-\Lambda\rho^3}\, \frac{A -R}{\rho}\right),\label{eq:App_Orr}\\
& \mathcal{O}_{\rho\vartheta} = \partial_\vartheta A + \rho^2\,\partial_\rho\Theta,\label{eq:App_Orth}\\
& \mathcal{O}_{\rho\varphi} = \partial_\varphi A + \rho^2\,\sin^2\vartheta\,\partial_\rho\Phi,\label{eq:App_Orph}\\
& \mathcal{O}_{\vartheta\vartheta} =2\rho^2\left(\partial_\vartheta\Theta + \frac{R}{\rho}\right),\label{eq:App_Othth}\\
& \mathcal{O}_{\vartheta\varphi} = \rho^2\left(\partial_\varphi\Theta + \sin^2\vartheta\,\partial_\vartheta\Phi\right),\label{eq:App_Othph}\\
& \mathcal{O}_{\varphi\varphi} = 2\rho^2\left( \partial_\varphi\Phi + \frac{R}{\rho}+ \Theta \cot\vartheta\right)\sin^2\vartheta ,\label{eq:App_Ophph}
\end{align}
\end{subequations}
where we have introduced the auxiliary function $A(\rho,\vartheta,\varphi) \equiv -\sqrt{\frac{2\mu(\rho)}{\rho}}\, T(\rho,\vartheta,\varphi) + \frac{\rho}{\rho-2\mu(\rho)}R(\rho,\vartheta,\varphi)$. So at a given order, we have to solve a system of coupled linear differential equations: the most general solution will be given by the sum of a homogeneous solution of $\mathcal{O}_{ij}\left[F^{(n)}_\mu\right] = 0$ and a particular solution.

\subsection{Homogeneous solution}

Let's first find a general solution of the system $\mathcal{O}_{ij}=0$, where $\mathcal{O}_{ij}$ is defined in \eqref{eq:App_Oij}. Eliminating all but $\Theta$-dependencies in the angular equations gives
\begin{equation}
\frac{1}{2\rho^2}\partial_\vartheta\left(\mathcal{O}_{\vartheta\vartheta}-\frac{\mathcal{O}_{\varphi\varphi}}{\sin^2\vartheta}\right) + \partial_\varphi\left(\frac{\mathcal{O}_{\vartheta\varphi}}{\rho^2\sin^2\vartheta}\right) =
\partial_\vartheta\left[\sin\vartheta\,\partial_\vartheta\left(\frac{\Theta}{\sin\vartheta}\right)\right] + \frac{1}{\sin^2\vartheta}\,\partial_\varphi^2\Theta =0.
\end{equation}
Imposing periodicity in $\varphi$ and regularity in $\vartheta$, the solution reads $\Theta = \Theta_s(\rho)\sin\vartheta +\Theta_c(\rho)\cos\vartheta\sin\left[\varphi-\varphi_0(\rho)\right]$. Plugging back in Eqs.~(\ref{eq:App_Othth}), (\ref{eq:App_Othph}) and~(\ref{eq:App_Ophph}) and solving them, it comes
$R = -\rho\,\Theta_s(\rho)\cos\vartheta+\rho\,\Theta_c(\rho)\sin\vartheta\sin\left[\varphi-\varphi_0(\rho)\right]$, and $\Phi = \Phi_0(\rho) + \Theta_c(\rho)\cos\left[\varphi-\varphi_0(\rho)\right]/\sin\vartheta$. Last but not least, $\partial_\vartheta\mathcal{O}_{\rho\varphi}-\partial_\varphi\mathcal{O}_{\rho\vartheta} = \rho^2\sin(2\vartheta)\Phi_0'$ forces $\Phi_0$ to be constant. Injecting those solutions in Eqs.~(\ref{eq:App_Orth}) and~(\ref{eq:App_Orph}) yields $A = A_0(\rho)+\rho^2\Theta_s'\cos\vartheta - \rho^2\sin\vartheta\partial_\rho\left(\Theta_c\sin\left[\varphi-\varphi_0(\rho)\right]\right)$. Finally Eq.~(\ref{eq:App_Orr}) gives
\begin{equation}
A_0'+\lambda\frac{A_0}{\rho} 
+ \left(\rho^2\Theta_s''+(2+\lambda)\rho\Theta_s'+\lambda\Theta_s \right)\cos\vartheta
-\left(\rho^2\tilde\Theta_c''+(2+\lambda)\rho\tilde\Theta_c'+\lambda\tilde\Theta_c \right)\sin\vartheta = 0,
\end{equation}
where we have shortened $\lambda = (3M+\Lambda\rho^3)/(6M-\Lambda\rho^3)$ and $\tilde\Theta_c = \Theta_c(\rho)\sin\left[\varphi-\varphi_0(\rho)\right]$.
This imposes $A_0 = -T_0\sqrt{2\mu/\rho} $, $\Theta_s = \left(\kappa_1+\kappa_2\int\dd u \sqrt{2\mu/u}\right)/\rho$ and $\Theta_c = 0$, where $T_0$, $\kappa_1$ and $\kappa_2$ are constants of integration. Let's note that in the $\Lambda=0$ limit, it simply comes $\Theta_s = \kappa_1/\rho+2\kappa_2\sqrt{2M/\rho}$. Turning back to the original variables (\emph{i.e.} expressing $T$ in terms of $A$ and $R$), it finally comes
\begin{subequations}\label{eq:App_coord_change_hom}
\begin{align}
& T_h = T_0 + \left\lbrace  \kappa_1\,\frac{\sqrt{2\rho\mu}}{\rho-2\mu}+ \kappa_2\,\left(\rho+\frac{\sqrt{2\rho\mu}}{\rho-2\mu}\int\!\!\dd u \sqrt{\frac{2\mu}{u}}\right)\right\rbrace \cos\vartheta, \\
& R_h = \left\lbrace  \kappa_1 + \kappa_2\, \int\!\!\dd u \sqrt{\frac{2\mu}{u}}\right\rbrace \cos\vartheta, \\
& \Theta_h = - \frac{1}{\rho}\left\lbrace \kappa_1 + \kappa_2 \,\int\!\!\dd u \sqrt{\frac{2\mu}{u}}\right\rbrace\sin\vartheta,\\
& \Phi_h = \Phi_0.
\end{align}
\end{subequations}
We can easily recognize that $T_0$ and $\Phi_0$ are respectively accounting for staticity and axisymmetry of the Schwarzschild-de Sitter metric. The two other constants are associated with the two remaining generators of the group of isometries of the Schwarzschild-de Sitter spacetime.

\subsection{At linear order}

At the linear order in $a$, the only non-vanishing source term is $\mathcal{S}^{(1)}_{\rho\varphi}$, leading to the equation
\begin{equation}
A^{(1)}_\varphi + \left[\rho^2\,\Phi^{(1)}_\rho -\left(\frac{2\mu}{\rho}\right)^{3/2}\frac{\rho}{\rho-2\mu}\right]\sin^2\vartheta = 0,
\end{equation}
which is easily solved by imposing $\Phi^{(1)}_p = \int^\rho\!\frac{\dd u}{u}\left(\frac{2\mu}{u}\right)^{3/2}\frac{1}{u-2\mu}.$
Together with the previously found homogeneous solution (\ref{eq:App_coord_change_hom}), it comes
\begin{subequations}\label{eq:App_coord_change_lin}
\begin{align}
& t = \tau + a T_0 + \int^\rho \! \dd u\, \frac{\sqrt{2Mu}}{u-2M} 
+ a \left\lbrace  \kappa_1\,\frac{\sqrt{2\rho\mu}}{\rho-2\mu}+ \kappa_2\,\left(\rho+\frac{\sqrt{2\rho\mu}}{\rho-2\mu}\int\!\!\dd u \sqrt{\frac{2\mu}{u}}\right)\right\rbrace \cos\vartheta
+ \mathcal{O}(a^2), \\
& r = \rho 
+ a\left\lbrace  \kappa_1 + \kappa_2\, \int\!\!\dd u \sqrt{\frac{2\mu}{u}}\right\rbrace\cos\vartheta+ \mathcal{O}(a^2), \\
& \theta = \vartheta - \frac{a}{\rho}\left\lbrace  \kappa_1 + \kappa_2\, \int\!\!\dd u \sqrt{\frac{2\mu}{u}}\right\rbrace\sin\vartheta + \mathcal{O}(a^2),\\
& \phi = \varphi + a \varphi_0  + a\int^\rho \!\frac{\dd u}{u} \left(\frac{2\mu}{u}\right)^{3/2}\frac{1}{u-2\mu} + \mathcal{O}(a^2).
\end{align}
\end{subequations}

Note that $\mathcal{J} = 1 + a\,\left(\kappa_1-\kappa_2\sqrt{2\mu\rho}+\kappa_2\int\!\!\dd u \sqrt{\frac{2\mu}{u}}\right)\frac{\cos\vartheta}{\rho}+\mathcal{O}(a^2)$ cannot vanish but in localized points.

\subsection{At second order}

At the second order, introducing $\mathcal{K} =\kappa_1+\kappa_2\int^\rho\!\!\dd u \sqrt{\frac{2\mu}{u}}$, the source term is slightly more complicated
\begin{subequations}\label{eq:App_Sij_quad}
\begin{align}
& \mathcal{S}^{(2)}_{\rho\rho} =\frac{\left(\rho+2\mu\right)\cos(2\vartheta)}{2\rho^3}-\frac{\rho^2-20\mu^2}{2\rho^3\left(\rho-2\mu\right)} -\frac{2M}{\rho^3}\frac{\rho^2+6\rho\mu-8\mu^2}{\left(\rho-2\mu\right)^2}\\
& \qquad \qquad \nonumber
-\left[\left(\frac{\left(3M-2\mu\right)^2\left(\rho^2-12\rho\mu-12\mu^2\right)}{2\mu\rho\left(\rho-2\mu\right)}+\rho-6\mu\right)\frac{\rho\,\cos^2\vartheta}{\left(\rho+2\mu\right)^2}-1\right]\frac{\mathcal{K}^2}{\rho^2}\\
& \qquad \qquad \nonumber
+2\frac{\sqrt{2\mu}}{\rho^{3/2}}\left(\frac{4\mu^2\left(\rho+2\mu\right)+3M\rho\left(\rho-6\mu\right)}{2\mu\left(\rho-2\mu\right)^2}\,\cos^2\vartheta-1\right)\,\kappa_2\mathcal{K}
+\frac{2\mu-\left(\rho-2\mu\right)\cos^2\vartheta}{\rho}\,\kappa_2^2,\\
& 
\mathcal{S}^{(2)}_{\rho\vartheta} = 
\left[\frac{\rho\left(\rho-6\mu\right)+3M\left(\rho+2\mu\right)}{\rho\left(\rho-2\mu\right)^2}\,\mathcal{K}^2 -\frac{3M}{\sqrt{2\rho\mu}}\,\kappa_2\mathcal{K}+\kappa_2^2\rho\right]\cos\vartheta\sin\vartheta,\\
& 
\mathcal{S}^{(2)}_{\rho\varphi} = 2\mu\,\left[\frac{2\mu\left(\rho+2\mu\right)-3M\left(3\rho-2\mu\right)}{\sqrt{2\mu}\,\rho^{3/2}}\,\mathcal{K}
+\left(\rho-2\mu\right)\kappa_2\right]\cos\vartheta\,\sin^2\vartheta,\\
& 
\mathcal{S}^{(2)}_{\vartheta\vartheta} = \left[1+\frac{\Lambda \rho^2}{3}-2\mathcal{K}^2\right]\cos^2\vartheta
+ \left[\left(\kappa_1+\left(\int\!\!\dd u\sqrt{\frac{2\mu}{u}}-\sqrt{2\rho\mu}\right)\kappa_2\right)^2-\kappa_2^2\,\rho^2\right]\sin^2\vartheta ,\\
& 
\mathcal{S}^{(2)}_{\vartheta\varphi} = \frac{2\mu}{\rho-2\mu} \,\left[\sqrt{\frac{2\mu}{\rho}}\,\mathcal{K}+(\rho+2\mu)\,\kappa_2\right] \sin^3\vartheta,\\
& 
\mathcal{S}^{(2)}_{\varphi\varphi} = \left[1 +\frac{\Lambda \rho^2}{3}+ \frac{2M}{\rho}\sin^2\vartheta -\left(1+\cos^2\vartheta \right)\mathcal{K}^2 \right] \sin^2\vartheta.
\end{align}
\end{subequations}

Let's first focus on the angular part. The combination $\partial_\vartheta\left(\mathcal{E}^{(2)}_{\vartheta\vartheta}-\frac{\mathcal{E}^{(2)}_{\varphi\varphi}}{\sin^2\vartheta}\right)+2\,\partial_\varphi\left(\frac{\mathcal{E}^{(2)}_{\vartheta\varphi}}{\sin^2\vartheta}\right)$ yields
\begin{equation}
\partial_\vartheta\left[\sin\vartheta\,\partial_\vartheta\left(\frac{\Theta^{(2)}_p}{\sin\vartheta}\right)\right] + \frac{\partial_\varphi^2\Theta^{(2)}_p}{\sin^2\vartheta}
=
\left[1+\frac{2M}{\rho}+\frac{\Lambda\rho^2}{3}-2\mathcal{K}^2+2\sqrt{2\rho\mu}\,\kappa_2\mathcal{K}+\kappa_2^2\,\rho\left(\rho-2\mu\right)\right]\frac{\sin(2\vartheta)}{2\rho^2}.
\end{equation}
This is notably solved by
\begin{equation}
 \Theta^{(2)}_p = -\left[1+\frac{2M}{\rho}+\frac{\Lambda\rho^2}{3}-2\mathcal{K}^2+2\sqrt{2\rho\mu}\,\kappa_2\mathcal{K}+\kappa_2^2\,\rho\left(\rho-2\mu\right)\right]\frac{\sin(2\vartheta)}{4\rho^2}\,,
\end{equation}
and thus, when injected back in $\mathcal{E}^{(2)}_{\vartheta\vartheta}-\frac{\mathcal{E}^{(2)}_{\varphi\varphi}}{\sin^2\vartheta}$ and $\mathcal{E}^{(2)}_{\vartheta\varphi}$, one obtains
\begin{equation}
\Phi^{(2)}_p = \left[\kappa_2+\sqrt{\frac{2\mu}{\rho}}\frac{\mathcal{K}}{\rho-2\mu}\right] \frac{2\mu\cos\vartheta}{\rho^2}\,.
\end{equation}
Then $\mathcal{E}^{(2)}_{\vartheta\vartheta}$ gives
\begin{equation}
 R^{(2)}_p = 
 \left(\frac{M}{\rho^2}+ \frac{\rho-2\mu}{2}\,\kappa_2^2+\sqrt\frac{2\mu}{\rho}\,\kappa_2\mathcal{K}\right)\cos^2\vartheta
 +\left(\frac{\mu-2M}{\rho^2}+\frac{\mathcal{K}^2-1}{2\rho}\right)\sin^2\vartheta\,.
\end{equation}

Turning to the radial part of the system, $\mathcal{E}^{(2)}_{\rho\varphi}$ gives
\begin{equation}
 A^{(2)}_p = \frac{6M\,\kappa_2\cos\vartheta\sin^2\vartheta}{2\rho}\, \varphi + f_2(\rho,\vartheta)\,,
\end{equation}
which, plugged into $\partial_\varphi\mathcal{E}^{(2)}_{\rho\vartheta}$ imposes that $\kappa_2 = 0$. $\mathcal{E}^{(2)}_{\rho\vartheta}$ then yields
\begin{equation}
 A^{(2)}_p =\left[\frac{3M+\rho}{\rho}-\frac{\rho^2-2\rho\mu+8\mu^2-3M\left(\rho+2\mu\right)}{\left(\rho+2\mu\right)^2}\right] \frac{\cos(2\vartheta)}{4\rho}\,.
\end{equation}
But it remains
\begin{equation}
\partial_\vartheta\mathcal{E}^{(2)}_{\rho\rho} = \frac{9M^2\sin(2\vartheta)}{2\rho^3\mu},
\end{equation}
that cannot vanish. Thus it is impossible to achieve a spatially flat slicing of the Kerr-de Sitter spacetime.

\section{Lengthy expressions of the second branch}\label{app:lengthy_expr}

In this Appendix, we present the lengthy expressions of Sec.~\ref{sec:cancel_Br2} in terms of the variable $z = \cos\theta$. So for example $\partial_z = -1/\sin\theta\,\partial_\theta$. Let's also recall our notation
\begin{equation}
\mathcal{A}_1 = r\left(1-\frac{2M}{r}\right)f_0''+\left(1-\frac{2M}{r}\right)^2\left(1-\frac{3M}{r}\right)(f_0')^3-\left(1-\frac{5M}{r}\right)f_0'.
\end{equation}

\paragraph{Second order:} The explicit form of the source term in the numerator of $\mathcal{C}^r_\phi$ (see Eq.~\eqref{eq:crphi_b2_a2}) is
\begin{equation}
\begin{aligned}
\mathcal{S}_{2} =\frac{1}{r-2M}&\left[ \frac{21M^2}{r^2}+r\left(1-\frac{2M}{r}\right)^3\left(1+\frac{3M}{r}\right)f_0'\,f_0''+\left(1-\frac{2M}{r}\right)^4\left(1+\frac{6M}{r}\right)(f_0')^4\right.\\
&\hspace{45ex}\left.-\left(1-\frac{2M}{r}\right)^2\left(1+\frac{4M}{r}+\frac{15M^2}{r^2}\right)(f_0')^2\right]\!,
\end{aligned}
\end{equation}
which is cancelled by the contribution of for $f_{2} = \bar{f}_2(r) + f_{2,c}(r)\cos\theta + \hat{f}_2(r)\cos^2\theta$, with
\begin{equation}
\hat{f}_2(r) = \frac{21M^2r^3 +(r-2M)^2\left[r^2(r-2M)(r+3M)f_0''+(r-2M)^2(r+6M)(f_0')^3-r(15M^2+4Mr+r^2)f_0'\right]f_0'}{2r^5(r-2M)\mathcal{A}_1}.
\end{equation}

\paragraph{Third order:} From $\mathcal{C}^r_r = 0$ one obtains the differential equation \eqref{eq:diffeq_f0} for $f_0$, solved by $\hat{f}_0$. Here we give this solution in terms of its first derivative,
\small
\begin{equation}
\hat{f}'_0 = \frac{C_{0,1}r^2 +\frac{7875M^2}{r^2} + \frac{34020M^3}{r^3} + \frac{51030 M^4}{r^4} + \frac{26244M^5}{r^5}}{\left(\frac{2M}{r}-1\right) \sqrt{\left(C_{0,1}r^2 + 1250 + \frac{6500M}{r} + \frac{14175M^2}{r^2} + \frac{14580M^3}{r^3} + \frac{5832M^4}{r^4}\right)\left(C_{0,1}r^2 + \frac{7875 M^2}{r^2} + \frac{34020M^3}{r^3} + \frac{51030M^4}{r^4} + \frac{26224M^5}{r^5}\right)}}\!.
\end{equation}
\normalsize
where $C_{0,1}$ is an integration constant. Next, considering $\mathcal{C}^r_\phi$ gives the differential equation \eqref{eq:diffeq_f3} for $f_3(r,\theta)$. The explicit form of the source term in it is
\begin{equation}
\begin{aligned}
\mathcal{S}_{3} =\frac{1}{1-(1-\frac{2M}{r})^2 \hat{f}_0'^2}&\left\{189\frac{M^2}{r^2}\left(1+\frac{3M}{r}\right)\bar{f}'_1 + \left(1-\frac{2M}{r}\right)^2\left[25 + \frac{91M}{r} + 9\frac{M^2}{r^2}\left(47+\frac{45M}{r}\right)\right]\hat{f}_0'^2\bar{f}'_1\right.\\
&\left.-3\left(1-\frac{2M}{r}\right)^4\left(1+\frac{3M}{r}\right)\left(25 + \frac{42M}{r}\right)\hat{f}_0'^4\bar{f}'_1 - \left(1-\frac{2M}{r}\right)^3\left(5+\frac{9M}{r}\right)^2\hat{f}_0'^2\bar{f}''_1 r \right\}\!,
\end{aligned}
\end{equation}

\paragraph{Fourth order:} From $\mathcal{C}^r_r = 0$ one obtains the differential equation \eqref{eq:diffeq_f1b} for $\bar{f}_1$, solved by $\hat{f}_1$. Here we give this solution in terms of its first derivative,
\small
\begin{equation}
\hat{f}'_1 = \frac{C_{1,1}\left(5 + \frac{9M}{r}\right)}{r \left(C_{0,1}r^2 + 1250 + \frac{6500M}{r} + \frac{14175M^2}{r^2} + \frac{14580M^3}{r^3} + \frac{5832M^4}{r^4}\right)^{3/2}\left(C_{0,1}r^2 + \frac{7875 M^2}{r^2} + \frac{34020M^3}{r^3} + \frac{51030M^4}{r^4} + \frac{26224M^5}{r^5}\right)^{1/2}},
\end{equation}
\normalsize
where $C_{1,1}$ and $C_{0,1}$ are integration constants.

At the fourth order there are also notably two source terms in $\mathcal{C}^r_\phi$, $\mathcal{S}_{4,1}$ and $\mathcal{S}_{4,3}$, the full equation 
\eqref{eq:diffeq_f4} being read as a differential equation for $f_4(r,\theta)$. Their explicit form (which does not vanish when replacing the functions $\hat{f}_0$ and $\hat{f}_1$) is
\begin{equation}
\begin{aligned}
\mathcal{S}_{4,3} =&-\frac{23814M}{r} - \frac{9M^2}{r^2}\left(1-\frac{2M}{r}\right)\left[475-\frac{247 M}{r} + \frac{162M^2}{r^2}\left(-21+\frac{38M}{r}\right)\right]\hat{f}'^2_0\\ &+\frac{M}{r}\left(1-\frac{2M}{r}\right)^4\left[1775 + \frac{14576M}{r} + \frac{81M^2}{r^2}\left(87 -\frac{314M}{r}+\frac{324M^2}{r^2}\right) \right]\hat{f}'^4_0\\
&+\left(1-\frac{2M}{r}\right)^6\left\{-625 -\frac{3235M}{r}+\frac{6M^2}{r^2}\left[-470+\frac{9M}{r}\left(146+\frac{135M}{r}\right)\right]\right\}\hat{f}'^6_0\\
&-\frac{54M}{r}\left(1-\frac{2M}{r}\right)^8\left(2+\frac{3M}{r}\right)\left(5+\frac{12M}{r}\right)\hat{f}'^8_0\,,
\end{aligned}
\end{equation}
\begin{equation}
\begin{aligned}
\mathcal{S}_{4,1} =\frac{2r^2\hat{f}_1' \hat{f}_0'^2\left(1-\frac{2M}{r}\right)^4\left(5+\frac{9M}{r}\right)^2}{\hat{f}_0'^2(1-\frac{2M}{r})^2 -1}\mathfrak{S}_1(r)+\frac{1}{3}\left[\mathfrak{S}_0(r) + \mathfrak{S}_2(r)\right],
\end{aligned}
\end{equation}
with
\begin{align}
\mathfrak{S}_1 =\,& \hat{f}_1' \left[5+ \frac{2M}{r}+ \frac{171M^2}{r^2}-5\left(1-\frac{2M}{r}\right)^2\left(5+ \frac{14M}{r} - \frac{9M^2}{r^2}\right)\hat{f}_0'^2\right]\nonumber\\& - r \hat{f}_1'' \left(5- \frac{M}{r}-\frac{18M^2}{r^2}\right)\left[1+\left(1-\frac{2M}{r}\right)\hat{f}_0'^2\right]\!,\\
\mathfrak{S}_0 =&\, \frac{23814M^4}{r^4} - 9 \hat{f}_0'^2\left(1-\frac{2M}{r}\right)\frac{M^2}{r^2}\left(1925 + \frac{11037M}{r} + \frac{13924 M^2}{r^2} - \frac{9936 M^3}{r^3} + \frac{14256 M^4}{r^4}\right)\nonumber\\
&-\hat{f}_0'^4\left(1-\frac{2M}{r}\right)^3\frac{M}{r}\left(1775 + \frac{4726M}{r}  - \frac{56665M^2}{r^2}  - \frac{72684M^3}{r^3}  + \frac{121824M^4}{r^4} \right)\nonumber\\
& +\hat{f}_0'^6\left(1-\frac{2M}{r}\right)^5\left(625 + \frac{3185M}{r}+ \frac{4870M^2}{r^2} + \frac{9804M^3}{r^3}+ \frac{36342M^4}{r^4} + \frac{26244M^5}{r^5}\right)\nonumber\\
& + \hat{f}_0'^8\left(1-\frac{2M}{r}\right)^8\frac{M}{r}\left(2 + \frac{3M}{r}\right)\left(5 + \frac{12M}{r}\right)\!,\\
\mathfrak{S}_2 =\,& 6\hat{f}_0'r^2\left(1 - \frac{2M}{r}\right)^2\left(5 + \frac{9M}{r}\right)\left\{r\bar{f}_2''\hat{f}_0'^2\left(1 - \frac{2M}{r}\right)^3\left(5 + \frac{9M}{r}\right)^2 + \bar{f}_2'\left[-\frac{189 M^2}{r^2}\left(1+\frac{3M}{r}\right) \right.\right.\nonumber\\
&\left.\left.-\hat{f}_0'^2\left(1-\frac{2M}{r}\right)^2\left(25 + \frac{91M}{r} + \frac{423M^2}{r^2}+ \frac{405M^3}{r^3}\right)+ 3\hat{f}_0'^4\left(1-\frac{2M}{r}\right)^4\left(25 + \frac{117M}{r} + \frac{126M^2}{r^2}\right)\right]\right\}\!.
\end{align}

\paragraph{Fifth order:} At this order it can be shown that a slicing $F(r,\theta,a)$ with $F(r,\theta,0) = f_0(r)$ cannot let the Cotton-York tensor $\mathcal{C}^i_j$ vanish since
\begin{equation}
\begin{aligned}
\frac{d^5}{d(\cos\theta)^5}\mathcal{C}_r^r = -\frac{7 a^5}{54 r^6 (5+\frac{9M}{r})^3 }\frac{\mathcal{N}_1(r)\mathcal{N}_2(r)}{\mathcal{D}_1(r)(\mathcal{D}_2(r))^{1/2}} + \mathcal{O}(a^6)\,,
\end{aligned}
\end{equation}
which we recall from \eqref{eq:Crr5_d5z}, does not identically vanish, this for any $C_{0,1}$. The explicit forms of the different terms are
\begin{equation}
\mathcal{N}_1 = C_{0,1}r^2+\frac{7875M^2}{r^2}+\frac{34020M^3}{r^3} + \frac{51030 M^4}{r^4} +\frac{26244 M^5}{r^5}\,,
\end{equation}
\begin{equation}
\mathcal{D}_1 =-C_{0,1}r^2+\frac{4375M}{r}+\frac{23625M^2}{r^2} +\frac{51030 M^3}{r^3} + \frac{51030 M^4}{r^4} + \frac{19683 M^5}{r^5}\,,
\end{equation}
\begin{equation}
\begin{aligned}
\mathcal{N}_2 =\, &25 C_{0,1}^2 r^4+10 C_{0,1}r^2\frac{M}{r} \left(9 C_{0,1}r^2-6175\right)-450 \frac{M^2}{r^2} \left(941 C_{0,1} r^2-331250\right)-270 \frac{M^3}{r^3} \left(4698 C_{0,1} r^2-5571875\right)\\
&-9477 \frac{M^4}{r^3} \left(288 C_{0,1} r^2-685625\right)- 7290\frac{M^5}{r^5}\left(513 C_{0,1} r^2-2167025\right)- 26244 \frac{ M^6}{r^6}\left(81 C_{0,1} r^2-895325\right)  \\
&+21379871430\frac{M^7}{r^7}+10979571060\frac{M^8}{r^8}+2448880128\frac{M^9}{r^9}\,,
\end{aligned}
\end{equation}
\begin{equation}
\begin{aligned}
\mathcal{D}_2 =\,& C_{0,1} r^2 \left(C_{0,1} r^2+1250\right)+6500 C_{0,1} \frac{M}{r} r^2+3150 \frac{M^2}{r^2} \left(7 C_{0,1} r^2+3125\right)+900 \frac{M^3}{r^3} \left(54 C_{0,1} r^2+104125\right)\\
&+243 \frac{M^4}{r^4} \left(234 C_{0,1} r^2+1631875\right)+2916 \frac{M^5}{r^5} \left(9 C_{0,1} r^2+329750\right)+1435874850 \frac{M^6}{r^6}+1314430740 \frac{M^7}{r^7}\\
&+680244480 \frac{M^8}{r^8}+153055008 \frac{M^9}{r^9}\,.
\end{aligned}
\end{equation}


\bibliography{CFKerr_biblio}

\end{document}